\documentclass[%
 reprint,
 amsmath,amssymb,
 aps,
floatfix,
]{revtex4-1}

\usepackage{graphicx}% Include figure files
\usepackage{dcolumn}% Align table columns on decimal point
\usepackage{bm}% bold math
\usepackage{placeins}
\usepackage{xcolor}
\usepackage{booktabs}
\usepackage{gensymb}
\usepackage[rightcaption]{sidecap}
\usepackage{newunicodechar}
\newunicodechar{ }{\,}

\begin{document}
% \preprint{APS/123-QED}

\title{Effect of annealing temperature on the structure and properties of co-sputtered Fe-Mn-Sn films near 2:1:1 ratio}

\author{Lance Griswold and Dipanjan Mazumdar}
\email{Corresponding author: dmazumdar@siu.edu}
 \affiliation{School of Physics and Applied Physics, Southern Illinois University, Carbondale, IL, 62901}
 
%\affiliation{

\date{\today}

% \keywords{Kagome, synthesis, Heusler}

\begin{abstract}
Research in recent years has focused on the thin-film synthesis of high-quality ternary alloys, identified for their tunable properties and potential in spintronics (e.g., Heusler alloys, Kagome magnets). In a previous study, we identified the conditions for stabilizing Fe$_2$MnSn, a Kagome magnet with a high Curie temperature and magnetic anisotropy. However, ternary phases such as Fe$_2$MnSn are challenging to synthesize and stabilize within a narrow temperature window, as binary and elemental phases can also form during the growth process. To highlight these observations, we investigated the thin film phases in the Fe-Mn-Sn system near the 2:1:1 ratio as a function of annealing temperature, ranging from 400 to 700\degree C. The elemental Fe, Mn, and Sn targets were pre-calibrated to a close to 2:1:1 ratio and co-sputtered at room temperature, followed by annealing. Two binary hexagonal structures, Fe$_3$Sn$_2$ and Fe$_5$Sn$_3$,  along with the elemental Fe phase, are stabilized between 400-550\degree C, but disappear at 580\degree C, where Fe$_2$MnSn is the only stable phase.  Elemental Mn phase starts to appear starting from 600\degree C, and becomes dominant by 750\degree C. Electrical, magnetic and magneto-optical properties are observed to correlate with the structural findings and the best properties are observed in the temperature range where Fe$_2$MnSn is the dominant phase. In general, our study highlights the difficulty in growing phase-pure ternary alloys such as Fe$_2$MnSn, which is very strongly based on precise temperature conditions. We also observed significant disordered growth below 100 nm for Fe$_2$MnSn, implying poor thickness scaling behavior.

\end{abstract}

\flushbottom

\maketitle
\thispagestyle{empty}

\section*{Introduction}

Interest in the synthesis and properties of hexagonal intermetallic magnets has surged in the area of spintronics as they show exotic topology-driven properties \cite{he_topological_2022,felser_topology_2022,yan_topological_2017,yin_topological_2022,blatova_topology_2024} coupled with a high magnetic ordering temperature and bulk magnetic anisotropy \cite{jami_tailoring_2025, faleev_heusler_2017,faleev_origin_2017}. Several Binary Fe-Sn and Mn-Sn alloys, such as Fe$_3$Sn$_2$ \cite{alikhan_intrinsic_2022, ren_plethora_2022, zhang_anomalous_2022,cheng_atomic_2022,khadka_anomalous_2020,lin_flatbands_2018} %IMPORTANT NOTE: Fe$_5$Sn$_3$ IS NOT KAGOME, IT IS SIDE-SHARING HEXAGONAL. SEE FIG 1c-1f for breakdown of structure.
, Fe$_3$Sn \cite{sales_ferromagnetism_2014, prodan_large_2023}, Mn$_3$Sn \cite{chen_anomalous_2021,higo_anomalous_2018,park_magnetic_2018} have received significant interest. Several studies have also explored ternary Fe-Mn-Sn system as they provide further tunability in terms of ordering temperature and magnetization \cite{dahal_electronic_2020, kratochvilova_fe2mnsn_2020,sapkota_synthesis_2025}. In particular, the properties of Mn$_{3-x}$Fe$_x$Sn (0$<$x$<$2) are interesting due to the desire to find materials with bulk magnetic perpendicular anisotropy for Magnetic Random Access Memory applications and tunable Curie temperature and magnetization \cite{prasad_material_2022,dieny_perpendicular_2017, faleev_heusler_2017,faleev_origin_2017,sato_perpendicular-anisotropy_2012,sato_magnetic_2017}. 
%In order to produce samples of sufficient quality, a process must be developed to encourage the phase desired. A phase diagram, if available, 
%paragraph addressing the processing portion of this work, with 5 citations minimum, mention the phase diagram found. Mention the many possibilities just from the binary phases,

%Kagome Magnets in general are explored due to exhibiting non-trivial magnetic textures (support Skyrmions) and spin frustration. In momentum space, they feature Dirac cones, which can enable Berry curvature, and flat bands, which are handy for generating strongly correlating electronic states. [Cheng][Linda Ye] Fe$_3$Sn$_2$ has been shown to grow at 400 C in vacuum better than 5x E-8 torr, with Pt and Ru seed layers  and is a high Anomalous Hall Effect Kagome magnet, with flat electronic bands.[Khadka][Cheng] Previously, Fe$_2$MnSn has been explored as a potential Huesler Alloy[Junaid Jami][Bishnu], but it has since been shown that the most stable structure is a hexagonal phase [Kratochilova]
%talk about binary FeSn phases, lead into Fe$_2$MnSn with High Anisotropy Avoid too much into our work until paragraph 3 or 4. Want a couple citations per sentence. At least 10 references before current work

 In our previous work, we synthesized Mn$_3$Sn, Mn$_2$FeSn, and Fe$_2$MnSn thin films by co-sputtering elemental targets at room temperature, followed by high-temperature annealing \cite{sapkota_synthesis_2025}. Fe$_2$MnSn, in particular, was recognized for its high Curie temperature of over 500K and reasonably high magnetic anisotropy, consistent with bulk reports \cite{dahal_electronic_2020,kratochvilova_fe2mnsn_2020}.
In particular, our investigations show that the synthesis of pure-phase Fe$_2$MnSn is limited to a narrow temperature range with strong binary Fe-Sn phases observed at lower annealing temperatures. 
In order to unravel such observations, we have conducted a structure-property investigation as a function of annealing temperature under the optimized 2:1:1 co-sputtered conditions for Fe-Mn-Sn. To upto 550$\degree$ C, the Fe$_2$MnSn phase is weak, and two binary Fe-Sn phases are observed that are identified as Fe$3$Sn$_2$, Fe$_5$Sn$_3$. By 580$\degree$ C, we obtain pure phase Fe$_2$MnSn, at which point the binary Fe-Sn binary phases are not observed. The Fe$_2$MnSn is stable until 750$\degree$ C, the highest temperature used in this study, but the presence of elemental Mn appears increasingly prominent beyond 600$\degree$ C. Broadly, our work highlights the stringent conditions under which ternary alloys such as Fe$_2$MnSn films can be synthesized.

\section*{Methods}

As part of a previous study, hundreds of samples were fabricated to obtain the growth and annealing conditions for hexagonal Mn$_{3-x}$Fe$_{x}$Sn (x=0,1,2) films \cite{sapkota_synthesis_2025} where a 580$^\circ$C annealing temperature was identified following room temperature deposition for optimal crystal quality. In this work, we used the growth conditions for Fe$_2$MnSn ( pre-calibrated 2:1:1 ratio), using the annealing temperature as the parameter of interest. DC magnetron co-sputtering of high-purity elemental targets (3N at minimum) was used to grow thin films on 10 mm square quartz substrates on a vacuum system with a base pressure better than $5\times10^{-8}$ Torr. Depositions were performed in an Argon atmosphere (6.6 mTorr) at room temperature.   While under vacuum,  samples were annealed at temperatures ranging between 400\degree - 750\degree C for 1 hour and analyzed for their structure and properties. To facilitate elemental peak identification, control samples from individual elemental targets were deposited, annealed and characterized separately, allowing direct comparison with peaks observed in co-sputtered thin films.
 \begin{figure*}[!ht]
    \centering
    \includegraphics[width=.875\linewidth]{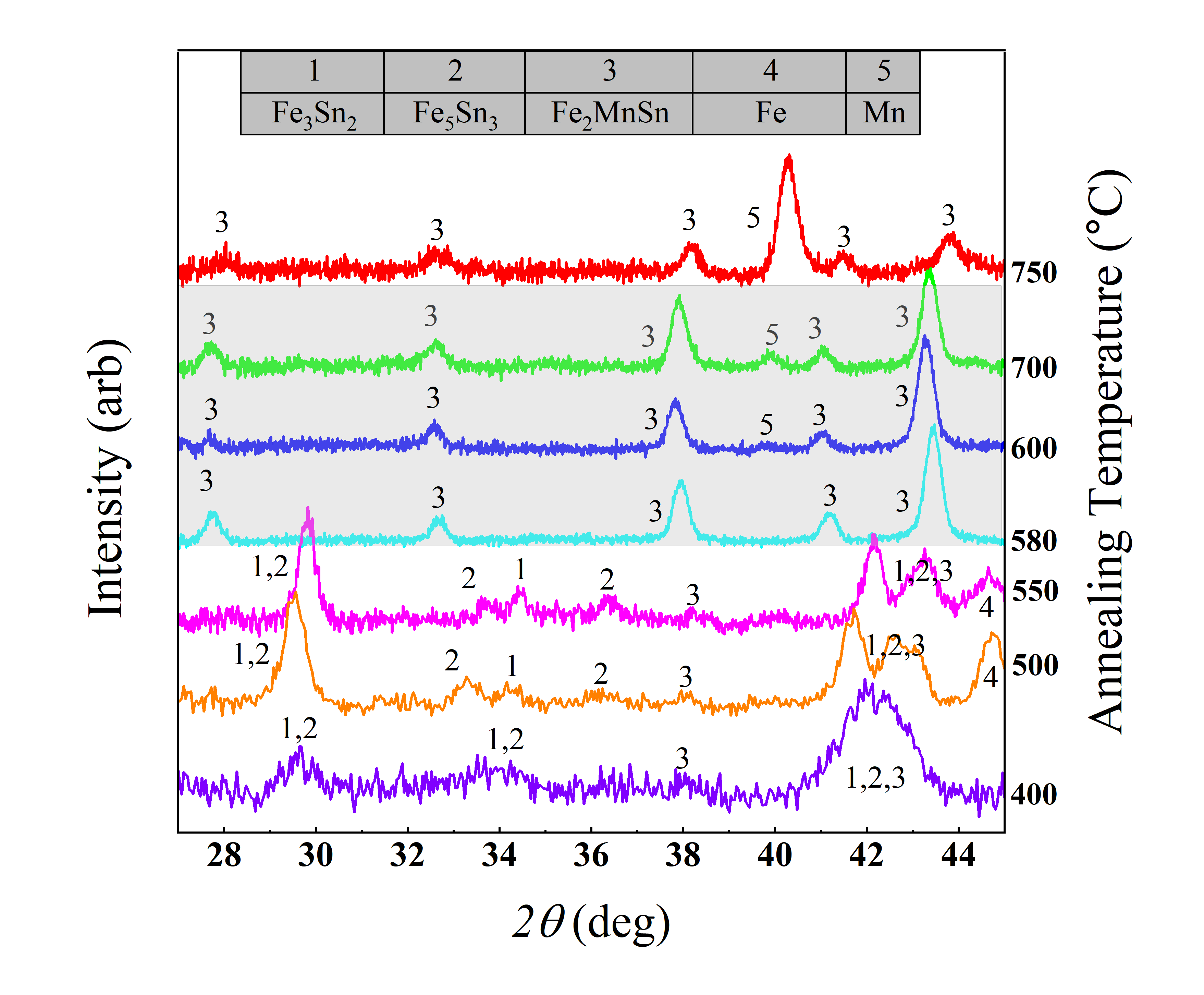}
    \caption{X-Ray Diffractions of samples annealed in the range from $400\degree-750\degree$C. At annealing temperatures below 580\degree C, the thin films exhibit a mixed phase, with a single hexagonal phase at 580\degree. As annealing temperature rises, a Mn isolation peak, becoming dominant at 750\degree C. Peaks are labeled to show the associated crystal structure}
    \label{XRDs}
 \end{figure*}
 
 X-Ray Diffractions (XRD) measurements were conducted to analyze the crystal structure using a Rigaku Smartlab Diffractometer with Cu- K$_{\alpha}$ rays ($\lambda$ =1.54 \AA) to conduct Bragg-Brentano and Parallel beam scans in the range 2$\theta$ from 27 to 45 degrees. Scans were analyzed for peak positions and reflections were compared with simulated diffraction data generated using VESTA \cite{momma_vesta_2011}. %Comparitive For elemental isolation peaks, a sample was grown under the same conditions using only one target and then scanned, providing peak positions near to those found in the diffractions of the annealing study samples. 
Electrical transport measurements were carried out using a standard four-probe inline configuration with a Signatone Pro4 probe station and a Keithley 2400 SourceMeter. Room-temperature magnetoresistance (MR) measurements were performed on selected samples using a custom-built MR measurement apparatus. Magneto-optical Kerr effect (MOKE) measurements were conducted in the longitudinal configuration using a custom-designed experimental setup. Surface morphology and roughness were further examined using atomic force microscopy (AFM).

%In this study, the Magneto-Optical Kerr Effect (MOKE) \cite{kerr_xliii_1877} is measured and applied, to determine some properties\cite{hamrle_magneto-optical_2001} of each sample in Fig. \ref{propertystack}. In particular, a lower coercivity material is considered a soft magnet,\cite{coisson_soft_2012} and can be used in a variety of devices, whereas an exceptionally high coercivity is considered magnetically hard\cite{jami_tailoring_2025}, and may be used  to make a permanent magnet.

\section*{Results and Discussion}
\begin{figure*}[h!]
    \centering
    \includegraphics[width=0.75\linewidth]{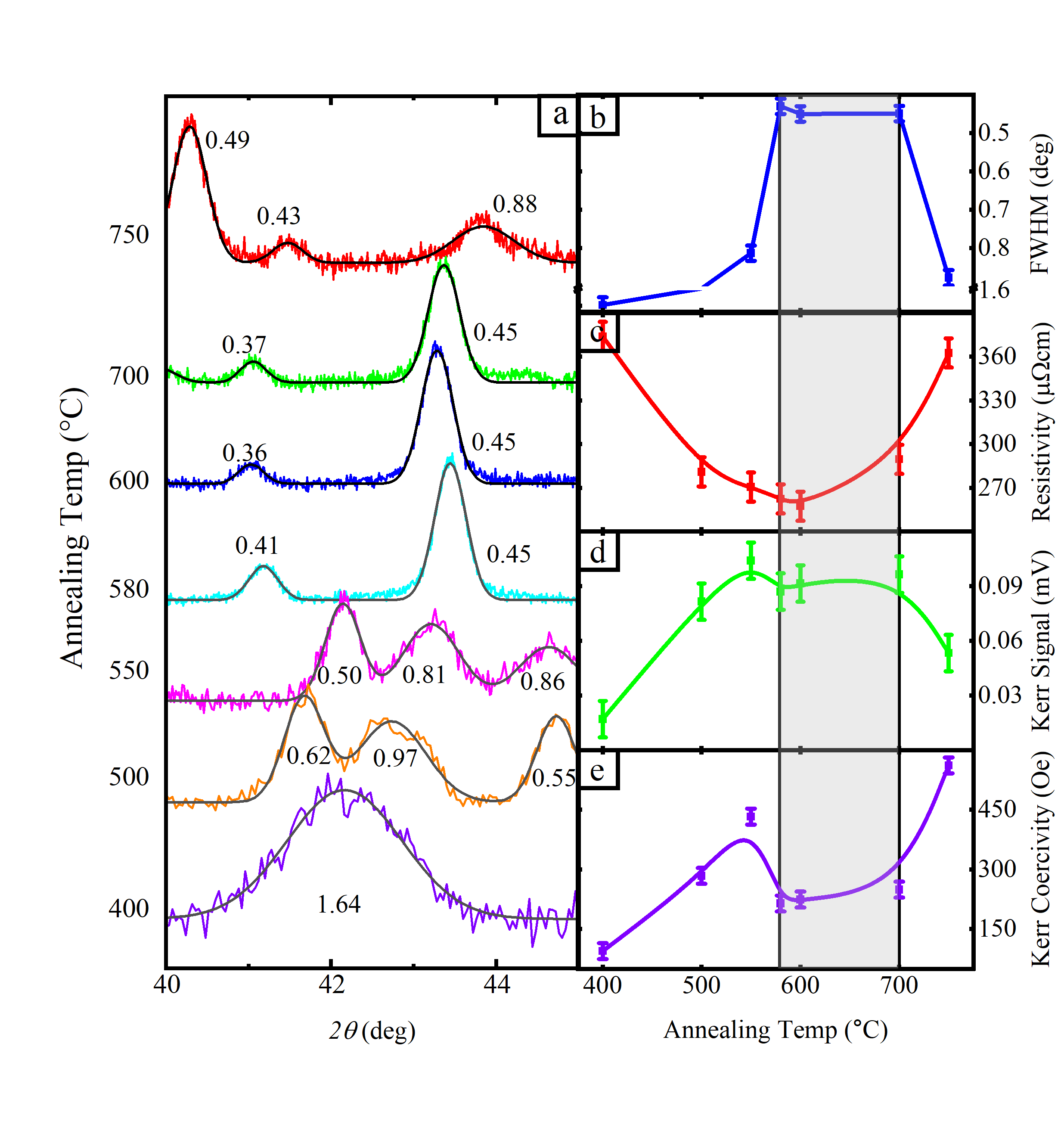}
    \caption{Structure-property relationship as a function of the annealing temperature. a) Data from Fig. \ref{XRDs} in the $2\theta$ range from 40\degree  to 45\degree  with line-shape analysis showing FWHM values. b) The FWHM value of (201) XRD peak of Fe$_2$MnSn or as appropriate for mixed phases. c) Resistivity as measured via four-probe line measurement showing a minimum in the 580-600\degree  range. d) The Kerr signal is maximized within the range optimized for Fe$_2$MnSn. e) Kerr Coercivity showing a minimum in the 580-600\degree range.}
    \label{propertystack}
\end{figure*}

Figure \ref{XRDs} shows the XRD data of films annealed at 400\degree, 500\degree,  550\degree, 580\degree, 600\degree, 700\degree and 750\degree C. XRD data between 27\degree-45\degree  is shown as the strongest peaks for all observed phases are well captured in this range. Up to 5 different phases were identified, as labeled in the figure, after an extensive search into possible elemental, binary, and ternary phases consistent with the overall 2:1:1 Fe:Mn:Sn ratio. 
%There is also a possibility that Mn$_3$Sn (201) diffraction peak appears in this range (~42\degree) \cite{sapkota_synthesis_2025}. But we are ruling it out as the nominal Mn:Sn is 1:1 ratio. Using a similar argument, Fe$_3$Sn may be less probable, but more likely than Mn$_3$Sn as Fe:Sn has a 2:1 ratio.  
For the sake of clarity, the data in 580-700\degree C range is highlighted, where Fe$_2$MnSn is the dominant, if not the only, phase. Only the sample annealed at 580\degree C exhibited a single-phase Fe$_2$MnSn, which was determined to be the optimum temperature in our previous report\cite{sapkota_synthesis_2025}. The impurity peaks are identified only 20 or 30 degrees outside of 580\degree C, highlighting some of the unique challenges of growth. Binary Fe-Sn phases and Fe appear very prominently at lower temperatures, whereas elemental Mn is the only observed impurity phase between 600-700 \degree C. At 750 \degree C, Mn completely dominates Fe$_2$MnSn.

We shall now discuss the evolution of the phases in the 400-550 $\degree$C range, which lies below the annealing temperature required to grow single-phase Fe$_2$MnSn. Line shape analysis of the peaks in the 40-45$\degree$ range is shown in Fig \ref{propertystack} (a) and will be discussed here as well. The sample annealed at 400$\degree$C marks the onset of crystallinity with several weak or broad diffraction peaks that match well with several binary Fe-Sn phases and Fe$_2$MnSn as indicated. In particular, the strongest peak observed at 42.1$\degree$ is also the broadest with a full-width-at-half-maxima (FWHM) of 1.64$\degree$, consistent with some of the strongest reflections for all binary Fe-Sn phases, including Fe$_3$Sn$_2$ (024) and (002), Fe$_5$Sn$_3$ (110) and (102), and Fe$_2$MnSn (201). All these reflections appear in the 42-43.5$\degree$ range. It is therefore not surprising that we observe the splitting of this peak going from 400$\degree$C to 500$\degree$C. We can identify three peaks based on our line-shape analysis for the 500$\degree$C sample as shown in Fig. \ref{propertystack}. The first is centered at 41.8$\degree$. We cannot assign a known binary phase reflection to this peak, implying that it may be a complicated off-stoichiometric Fe-Sn phase or involve Mn as well. The second peak at 42.8$\degree$  matches Fe$_5$Sn$_3$ ((102) and (110)), Fe$_3$Sn$_2$(024) and Fe$_2$MnSn (201). We also identify a bcc-Fe (110) peak at 44.7$\degree$. There is no clear evidence of Mn phases in this range.

\begin{figure}[h!]
     \centering
     \includegraphics[width=1\linewidth]{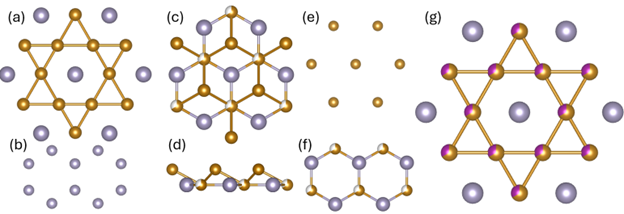}
     \caption{Crystal Structure of the phases present: (a) Fe$_3$Sn Kagome layer of Fe$_3$Sn$_2$, (b) Sn honeycomb lattice layer of Fe$_3$Sn$_2$, (c) frontal view of repeating Fe$_5$Sn$_3$ layers, (d) profile view of the same, (e) Hexagonal lattice of Fe-I, (f) Honeycomb Lattice formed of Sn and Fe-II, which has an occupancy of 0.67, and (g) Fe$_2$MnSn Kagome, formed with a shared occupancy between Fe at 0.67 and Mn at 0.33}
     \label{Xtals}
 \end{figure}

 The other most significant change at 500$\degree$  is observed around 29.5$\degree$ where the peak sharpens significantly from 400\degree to be the strongest peak in the entire XRD spectrum. Based on our analysis, this peak is the strongest signature of the Fe-Sn phases, and we assign it to Fe$_3$Sn$_2$ (015), Fe$_5$Sn$_3$ (101). Fe$_3$Sn$_2$, Fe$_5$Sn$_3$ have similar but varying hexagonal structures as shown in Fig. \ref{Xtals}. Fe$_3$Sn$_2$ belongs to the space group R$\overline{3}$m with lattice parameters a = b = 5.34 \AA~ and c = 19.79 \AA~. There are two stacked layers present, the first of which (a) is the Fe$_3$Sn Kagome structure, and the second (b) is a honeycomb lattice of monolayer Sn (Stanene) \cite{khadka_anomalous_2020, dally_isotropic_2021}. Fe$_5$Sn$_3$ (c) and (d), has space group P 63/ m m c with lattice parameters a = b = 4.224 \AA~ and c = 5.222 \AA~. It has two alternating layers, the first, known as Fe-I, is a hexagonal lattice, and the second is a honeycomb lattice of Sn and Fe-II, which has an occupancy of 0.67. Each Fe-I bonds with up to three Fe-II’s, depending on how many are vacant, and the same is true for each Sn atom. \cite{ren_plethora_2022}.

 From 500$\degree$ to 550\degree, all the peaks shift to higher values of $2\theta$, indicating lattice contraction without significant changes in relative intensity. For example, the 29.5$\degree$ peak is shifted to 29.85$\degree$, which conforms very well to the relaxed bulk lattice structure. Overall, the 400-550$\degree$ data demonstrate that this temperature range promotes the growth of the binary Fe-Sn phases, and is it not possible to grow pure phase Fe$_2$MnSn in this temperature window using our growth technique.

 %This shows potential for greater coherence in the crystal structure at higher annealing temperatures, as well as an opportunity to produce highly anisotropic materials if needed.
%Splitting the XRD analysis into FeSn focus and Fe$_2$MnSn focus

At 580\degree, the crystal phases snaps into a single hexagonal Fe$_2$MnSn phase with no evidence of the Fe-Sn binary or any other impurity phases. Fe$_2$MnSn is a bilayer kagome and belongs to the P 63/ m m c space group with lattice parameters a = b = 5.49 \AA~ and c = 4.38 \AA~. Fig \ref{Xtals} (c)  depicts one of the Kagome layers formed with a shared occupancy between Fe at 0.67 and Mn at 0.33 within a hexagonal lattice of Sn. As observed very clearly, the strongest Fe$_2$MnSn  peak appears at 43.4$\degree$ with a FWHM of 0.45$\degree$ and is indexed as the (201) peak. The other peaks appear at 27.8\degree (110), 32.6\degree (101), 38\degree (200), 41.2 (002). At 600 \degree C, we observe trace amounts of Mn as the impurity, but Fe$_2$MnSn is overwhelmingly the dominant phase until 700\degree, maintaining the FWHM of the (201) peak throughout.  

 %becomes the dominant phase by 750\degree. structure. The third group, covering temperatures of 700\degree and above, shows the appearance and dominance of an isolation peak of Manganese, marking an additional phase, further bringing issue to sample production. The sample annealed at 700\degree  can also be associated and compared to that at 600\degree to show the dominance of isolated Mn in the sample. 

%	Drop lines are placed along the temperature axis at 550°C and 700°C to highlight the range that Fe2MnSn is dominant.
%Kagome in Orange. 	Grey box added to highlight range dominated by Fe2MnSn
%	b) Resistivity as measured via four-probe inline measurement. In the grey ’optimal’ range, resistivity is minimized.
%	c) The Kerr signal is maximized within the range optimized for 	Fe2MnSn, demonstrating a more 	optically tunable phase.
%	d) Kerr Coercivity is an indicator of how permanent a magnetized 	state is, akin to resisting change in the magnetic moment. After the 	structure stabilized into Fe2MnSn, the Kerr Coercivity remains 	consistent, between 200 and 250 Oe, until the structure is dominated by Mn isolation peak.

In Fig. \ref{propertystack}, we also demonstrate structure-property relationships for the annealed samples. The structure is quantified by the FWHM of  Fe$_2$MnSn (201) reflection near 43.3\degree. In the low temperature range, we have used the FWHM of the appropriate peak in the 42-44$\degree$ range associated with Fe$_2$MnSn. In Fig. \ref{propertystack}b, we clearly observe the lowest FWHM (0.45\degree) in the 580-700$\degree$ range, which corresponds to almost phase pure Fe$_2$MnSn.
The limitation of this approach is that it does not capture the presence of the impurity phases that also influence the properties we report here.

Tracking this observation to the electronic properties, in Fig. \ref{propertystack}(c), we observe a clear resistivity minimum of close to 250 $\mu\Omega$cm in the the 580-600$\degree$C window that corresponds very well to the optimum growth range identified for Fe$_2$MnSn. We observe the resistivity value increase steadily to near 300 $\mu\Omega$cm by 700$\degree$C (and 500$\degree$C on the low temperature side), and sharply to over 360 $\mu\Omega$cm by 750$\degree$C / 400$\degree$ C . The resistivity data, therefore, tracks very sharply to the Fe$_2$MnSn phase, with impurity phases and disorder leading to enhance scattering and higher resistivity, as a result.

 \begin{figure*}[t]
    \centering
    \includegraphics[width=0.75\linewidth]{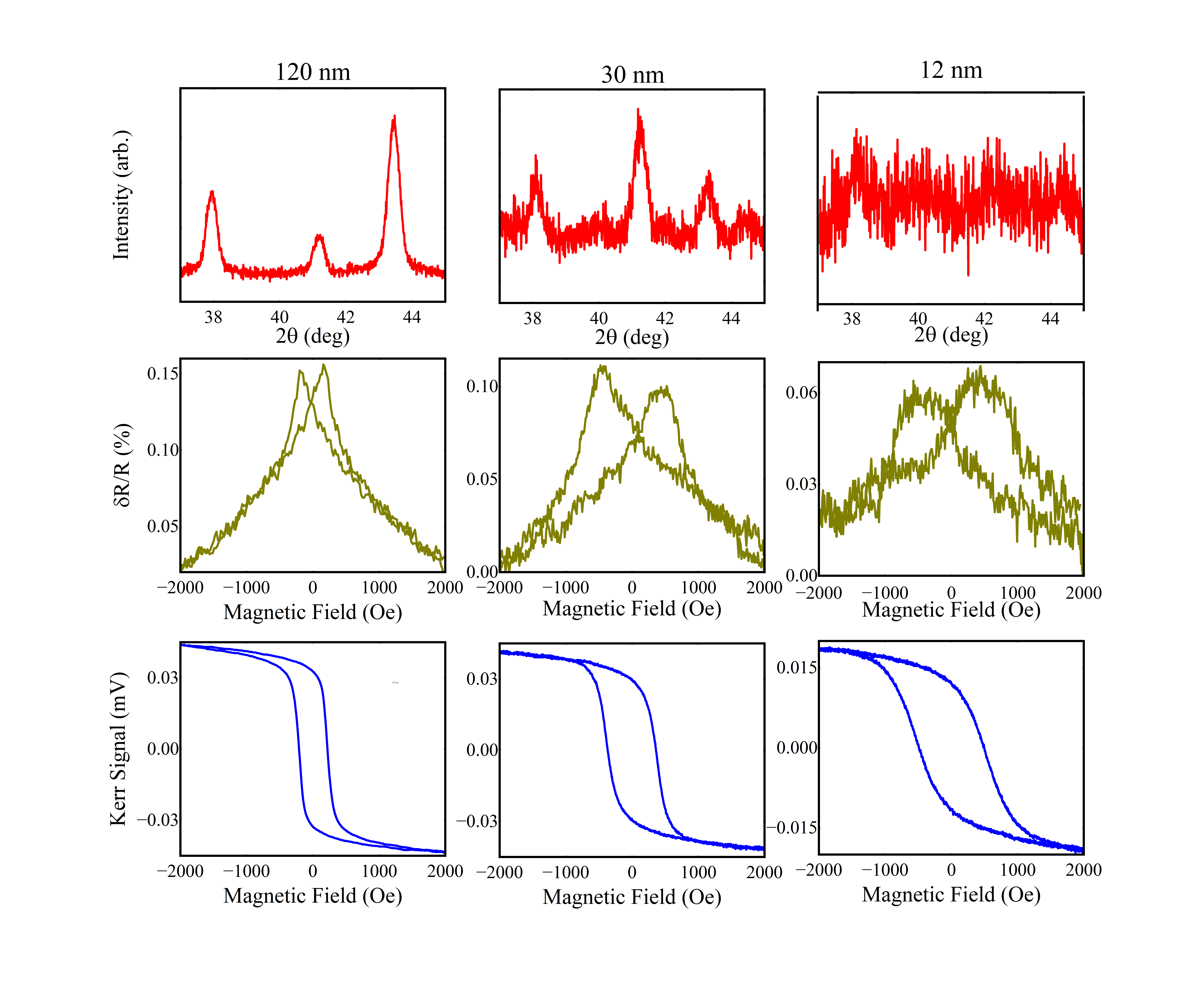}
    \caption{Thickness dependency study, showing the degradation of Fe$_2$MnSn samples grown on quartz as thickness decreases. Apart from structure (first row), the measured properties are magnetoresistance (second row) and Kerr signal (third row). Columns correspond to thicknesses as indicated}
    \label{thickness}
\end{figure*}

On the magnetic property side, we also find a similar, but somewhat weaker, trend within the 580-700$\degree$C window as measured by the Kerr signal (Fig. \ref{propertystack}(d)), which is proportional to the magnetization, and the coercivity (Fig. \ref{propertystack}(e)), that is related to the crystal and microstructure. Probably the most striking finding is the drop in coercivity, and the magnetization to a smaller extent, going from 550$\degree$ to 580$\degree$. This captures elegantly how quickly the multi-phase system collapses to a single-phase entity (within 30 degrees). Both magnetization and coercivity remain fairly constant over the 580-700\degree range, beyond which the coercivity(magnetization) increases(decreases) due to the formation of disordered antiferromagnetic Mn phase.
On the low-temperature side, both magnetization and coercivity increase steadily starting from 400$\degree$ and reaches a maximum at 550$\degree$, beyond which is drops sharply as mentioned earlier.

%shows the intensities of large peaks from differing phases being tracked as annealing temperature changes. A drop line has been added to emphasize the abrupt change in which crystal structures are dominant in the system. This is used to determine the ideal annealing temperature to maximize the presence of Fe$_2$MnSn within the sample. While the exact annealing time and temperature can vary while optimizing other factors such as atmospheric pressure during sputtering, this gives a baseline to work with. This figure also shows three properties of the annealed samples which are useful when determining the quality and cleanliness of the crystal structures within the sample, such as a relatively low resistivity indicating a more regular, higher quality crystal structure. Recalling the previously mentioned phases from the XRD analysis, the crystal structure snaps into place at 580\degree C. With the Fe$_3$Sn$_2$ and Fe$_5$Sn$_3$ becoming unstable at 550\degree c. The drop line at 700\degree C indicates the point where the unknown phase begins to become a more noticeable presence, though it initially appears as early as 600 degrees. Connecting resistivity and Kerr effect, they appear to have an inverse correlation, suggesting Kerr effect can be optimized similarly to conductivity.

We also investigated, as shown in Fig. \ref{thickness}, the effect of reducing thickness on the structural and magnetic properties of Fe$_2$MnSn. The thicknesses were chosen relative to the 100 nm sample (considered to be be the reference), and only two other thicknesses (30 and 12 nm) were considered here as the properties degraded very rapidly with reducing thickness. For clarity, we show the XRD scans in the 37-45$\degree$ range.  We observed weak XRD for the 30 nm film with a relatively strong (002) reflection, indicating a textured c-oriented growth at lower thickness. No observable XRD was detected for the 12 nm film. Anisotropic magnetoresistance  and Ker effect measurements are consistent with the structural trends. The AMR ratio drops from 0.15\% for the 100 nm film to 0.1 \% at 30 nm and 0.06 \% for the 12 nm film. 
The coercivity in AMR and Kerr measurements increases significantly with reducing thickness, again consistent with poor microstructure and crystal quality.  We also observe reduction in the the squareness of the hysteresis loops degrades sharply. The Kerr signal, which is proportional to magnetization, is significantly reduced going from 30 to 12 nm, implying that both magnetic and magnetotransport properties show poor scaling behavior in Fe$_2$MnSn films.

\section*{Conclusions}
	In conclusion, we have demonstrated the challenges in growing pure-phase ternary alloys such as Fe$_2$MnSn. In this study, our growth approach involved a room-temperature deposition of elemental Fe,Mn, and Sn in a 2:1:1 ratio followed by high-temperature annealing of between 400-750$\degree$C. We highlight, in particular, that low-temperature annealing, particularly in the 400-550$\degree$ range, promotes the growth of binary Fe-Sn phases, but not at 580$\degree$ or above, where Fe$_2$MnSn is the dominant phase until 700$\degree$C. Several properties such as transport and magnetism also correlate very well with the observed structural properties. Additionally, we also reveal that Fe$_2$MnSn shows poor scaling behavior with thickness.

\section{ACKNOWLEDGEMENTS}
The authors wish to acknowledge support from the NSF CAREER grant (ECCS 1846829) and Ian Logue, Thayne Dean and Noah Ross for their assistance during the growth of samples.
    
\bibliography{BIBLIO}

@article{alikhan_intrinsic_2022,
	title = {Intrinsic anomalous {Hall} effect in thin films of topological kagome ferromagnet {Fe} 3 {Sn} 2},
	volume = {14},
	url = {https://pubs.rsc.org/en/content/articlelanding/2022/nr/d2nr00443g},
	doi = {10.1039/D2NR00443G},
	number = {23},
	urldate = {2026-01-21},
	journal = {Nanoscale},
	publisher = {Royal Society of Chemistry},
	author = {Ali Khan, Kacho Imtiyaz and Singh Yadav, Ram and Bangar, Himanshu and Kumar, Akash and Chowdhury, Niru and Kumar Muduli, Prasanta and Kishor Muduli, Pranaba},
	year = {2022},
	pages = {8484--8492},
}

@article{dahal_electronic_2020,
	title = {Electronic, magnetic, and structural properties of {Fe2MnSn} {Heusler} alloy},
	volume = {10},
	issn = {2158-3226},
	url = {https://doi.org/10.1063/1.5127671},
	doi = {10.1063/1.5127671},
	number = {1},
	urldate = {2026-01-21},
	journal = {AIP Advances},
	author = {Dahal, Bishnu and Al Maruf, Abdullah and Prophet, Sam and Huh, Yung and Lukashev, Pavel V. and Kharel, Parashu},
	month = jan,
	year = {2020},
	pages = {015118},
}

@article{kratochvilova_fe2mnsn_2020,
	title = {{Fe2MnSn} – {Experimental} quest for predicted {Heusler} alloy},
	volume = {501},
	issn = {0304-8853},
	url = {https://www.sciencedirect.com/science/article/pii/S0304885319333128},
	doi = {10.1016/j.jmmm.2020.166426},
	urldate = {2026-01-21},
	journal = {Journal of Magnetism and Magnetic Materials},
	author = {Kratochvílová, M. and Král, D. and Dušek, M. and Valenta, J. and Colman, R. H. and Heczko, O. and Veis, M.},
	month = may,
	year = {2020},
	pages = {166426},
}

@misc{jami_tailoring_2025,
	title = {Tailoring hard magnetic properties of {Fe2MnSn} {Heusler} alloy via interstitial modification: {A} first-principles approach},
	shorttitle = {Tailoring hard magnetic properties of {Fe2MnSn} {Heusler} alloy via interstitial modification},
	url = {http://arxiv.org/abs/2507.01832},
	doi = {10.48550/arXiv.2507.01832},
	urldate = {2026-01-21},
	publisher = {arXiv},
	author = {Jami, Junaid and Pathak, Rohit and Venkataramani, N. and Suresh, K. G. and Bhattacharya, Amrita},
	month = jul,
	year = {2025},
	note = {arXiv:2507.01832 [cond-mat]},
}

@article{sapkota_synthesis_2025,
	title = {Synthesis and properties of stoichiometric {Mn3}-{xFexSn} thin films},
	volume = {9},
	issn = {29498228},
	url = {https://linkinghub.elsevier.com/retrieve/pii/S2949822825007294},
	doi = {10.1016/j.nxmate.2025.101211},
	urldate = {2025-10-10},
	journal = {Next Materials},
	author = {Sapkota, Yub Raj and Llamazares, J.L.Sánchez and Hofer, Stephen and Wetzel, Duston and Stiwinter, Kenneth and Griswold, Lance and Mazumdar, Dipanjan},
	month = oct,
	year = {2025},
	pages = {101211},
}

@article{he_topological_2022,
	title = {Topological spintronics and magnetoelectronics},
	volume = {21},
	copyright = {2021 Springer Nature Limited},
	issn = {1476-4660},
	url = {https://www.nature.com/articles/s41563-021-01138-5},
	doi = {10.1038/s41563-021-01138-5},
	number = {1},
	urldate = {2025-09-15},
	journal = {Nature Materials},
	publisher = {Nature Publishing Group},
	author = {He, Qing Lin and Hughes, Taylor L. and Armitage, N. Peter and Tokura, Yoshinori and Wang, Kang L.},
	month = jan,
	year = {2022},
	pages = {15--23},
}

@article{yin_topological_2022,
	title = {Topological kagome magnets and superconductors},
	volume = {612},
	issn = {0028-0836, 1476-4687},
	url = {https://www.nature.com/articles/s41586-022-05516-0},
	doi = {10.1038/s41586-022-05516-0},
	language = {en},
	number = {7941},
	urldate = {2025-10-22},
	journal = {Nature},
	author = {Yin, Jia-Xin and Lian, Biao and Hasan, M. Zahid},
	month = dec,
	year = {2022},
	pages = {647--657},
}

@article{dally_isotropic_2021,
	title = {Isotropic {Nature} of the {Metallic} {Kagome} {Ferromagnet} {Fe3Sn2} at {High} {Temperatures}},
	volume = {11},
	issn = {2073-4352},
	url = {https://www.mdpi.com/2073-4352/11/3/307},
	doi = {10.3390/cryst11030307},
	language = {en},
	number = {3},
	urldate = {2025-10-22},
	journal = {Crystals},
	author = {Dally, Rebecca L. and Phelan, Daniel and Bishop, Nicholas and Ghimire, Nirmal J. and Lynn, Jeffrey W.},
	month = mar,
	year = {2021},
	pages = {307},
}

@article{ren_plethora_2022,
	title = {Plethora of tunable {Weyl} fermions in kagome magnet {Fe3Sn2} thin films},
	volume = {7},
	issn = {2397-4648},
	url = {https://www.nature.com/articles/s41535-022-00521-y},
	doi = {10.1038/s41535-022-00521-y},
	language = {en},
	number = {1},
	urldate = {2025-10-22},
	journal = {npj Quantum Materials},
	author = {Ren, Zheng and Li, Hong and Sharma, Shrinkhala and Bhattarai, Dipak and Zhao, He and Rachmilowitz, Bryan and Bahrami, Faranak and Tafti, Fazel and Fang, Shiang and Ghimire, Madhav Prasad and Wang, Ziqiang and Zeljkovic, Ilija},
	month = nov,
	year = {2022},
	pages = {109},
}

@article{khadka_anomalous_2020,
	title = {Anomalous {Hall} and {Nernst} effects in epitaxial films of topological kagome magnet {Fe} 3 {Sn} 2},
	volume = {4},
	issn = {2475-9953},
	url = {https://link.aps.org/doi/10.1103/PhysRevMaterials.4.084203},
	doi = {10.1103/PhysRevMaterials.4.084203},
	language = {en},
	number = {8},
	urldate = {2025-10-22},
	journal = {Physical Review Materials},
	author = {Khadka, Durga and Thapaliya, T. R. and Hurtado Parra, Sebastian and Wen, Jiajia and Need, Ryan and Kikkawa, James M. and Huang, S. X.},
	month = aug,
	year = {2020},
	pages = {084203},
}

@article{lin_flatbands_2018,
	title = {Flatbands and {Emergent} {Ferromagnetic} {Ordering} in {Fe} 3 {Sn} 2 {Kagome} {Lattices}},
	volume = {121},
	issn = {0031-9007, 1079-7114},
	url = {https://link.aps.org/doi/10.1103/PhysRevLett.121.096401},
	doi = {10.1103/PhysRevLett.121.096401},
	language = {en},
	number = {9},
	urldate = {2025-10-22},
	journal = {Physical Review Letters},
	author = {Lin, Zhiyong and Choi, Jin-Ho and Zhang, Qiang and Qin, Wei and Yi, Seho and Wang, Pengdong and Li, Lin and Wang, Yifan and Zhang, Hui and Sun, Zhe and Wei, Laiming and Zhang, Shengbai and Guo, Tengfei and Lu, Qingyou and Cho, Jun-Hyung and Zeng, Changgan and Zhang, Zhenyu},
	month = aug,
	year = {2018},
	pages = {096401},
}

@article{cheng_atomic_2022,
	title = {Atomic layer epitaxy of kagome magnet {Fe3Sn2} and {Sn}-modulated heterostructures},
	volume = {10},
	issn = {2166-532X},
	url = {https://pubs.aip.org/apm/article/10/6/061112/2835067/Atomic-layer-epitaxy-of-kagome-magnet-Fe3Sn2-and},
	doi = {10.1063/5.0094257},
	language = {en},
	number = {6},
	urldate = {2025-10-22},
	journal = {APL Materials},
	author = {Cheng, Shuyu and Wang, Binbin and Lyalin, Igor and Bagués, Núria and Bishop, Alexander J. and McComb, David W. and Kawakami, Roland K.},
	month = jun,
	year = {2022},
	pages = {061112},
}

@article{zhang_anomalous_2022,
	title = {Anomalous and topological {Hall} effects of ferromagnetic {Fe3Sn2} epitaxial films with kagome lattice},
	volume = {120},
	issn = {0003-6951, 1077-3118},
	url = {https://pubs.aip.org/apl/article/120/23/232401/2833788/Anomalous-and-topological-Hall-effects-of},
	doi = {10.1063/5.0096144},
	language = {en},
	number = {23},
	urldate = {2025-10-21},
	journal = {Applied Physics Letters},
	author = {Zhang, Dongyao and Hou, Zhipeng and Mi, Wenbo},
	month = jun,
	year = {2022},
	pages = {232401},
}

@article{blatova_topology_2024,
	title = {Topology of {Intermetallic} {Crystals}: {Classification}, {Uniformity}, and {Transitions}},
	volume = {63},
	copyright = {https://doi.org/10.15223/policy-029},
	issn = {0020-1669, 1520-510X},
	shorttitle = {Topology of {Intermetallic} {Crystals}},
	url = {https://pubs.acs.org/doi/10.1021/acs.inorgchem.4c03033},
	doi = {10.1021/acs.inorgchem.4c03033},
	language = {en},
	number = {38},
	urldate = {2025-09-12},
	journal = {Inorganic Chemistry},
	author = {Blatova, Olga A. and Slavnov, Tikhon D. and Afanasieva, Anastasiya D. and Grebennikov, Alexei M. and Blatov, Vladislav A.},
	month = sep,
	year = {2024},
	pages = {17881--17890},
}

@article{yan_topological_2017,
	title = {Topological {Materials}: {Weyl} {Semimetals}},
	volume = {8},
	issn = {1947-5454, 1947-5462},
	shorttitle = {Topological {Materials}},
	url = {https://www.annualreviews.org/doi/10.1146/annurev-conmatphys-031016-025458},
	doi = {10.1146/annurev-conmatphys-031016-025458},
	language = {en},
	number = {1},
	urldate = {2025-08-29},
	journal = {Annual Review of Condensed Matter Physics},
	author = {Yan, Binghai and Felser, Claudia},
	month = mar,
	year = {2017},
	pages = {337--354},
}

@article{higo_anomalous_2018,
	title = {Anomalous {Hall} effect in thin films of the {Weyl} antiferromagnet {Mn3Sn}},
	volume = {113},
	issn = {0003-6951, 1077-3118},
	url = {https://pubs.aip.org/apl/article/113/20/202402/594316/Anomalous-Hall-effect-in-thin-films-of-the-Weyl},
	doi = {10.1063/1.5064697},
	language = {en},
	number = {20},
	urldate = {2025-05-16},
	journal = {Applied Physics Letters},
	author = {Higo, Tomoya and Qu, Danru and Li, Yufan and Chien, C. L. and Otani, Yoshichika and Nakatsuji, Satoru},
	month = nov,
	year = {2018},
	pages = {202402},
}

@article{chen_anomalous_2021,
	title = {Anomalous transport due to {Weyl} fermions in the chiral antiferromagnets {Mn3X}, {X} = {Sn}, {Ge}},
	volume = {12},
	issn = {2041-1723},
	url = {https://doi.org/10.1038/s41467-020-20838-1},
	doi = {10.1038/s41467-020-20838-1},
	number = {1},
	journal = {Nature Communications},
	author = {Chen, Taishi and Tomita, Takahiro and Minami, Susumu and Fu, Mingxuan and Koretsune, Takashi and Kitatani, Motoharu and Muhammad, Ikhlas and Nishio-Hamane, Daisuke and Ishii, Rieko and Ishii, Fumiyuki and Arita, Ryotaro and Nakatsuji, Satoru},
	month = jan,
	year = {2021},
	pages = {572},
}

@article{momma_vesta_2011,
	title = {\textit{{VESTA} 3} for three-dimensional visualization of crystal, volumetric and morphology data},
	volume = {44},
	issn = {0021-8898},
	url = {https://journals.iucr.org/paper?S0021889811038970},
	doi = {10.1107/S0021889811038970},
	number = {6},
	urldate = {2025-05-09},
	journal = {Journal of Applied Crystallography},
	author = {Momma, Koichi and Izumi, Fujio},
	month = dec,
	year = {2011},
	pages = {1272--1276},
}

@article{felser_topology_2022,
	title = {Topology, skyrmions, and {Heusler} compounds},
	volume = {47},
	issn = {0883-7694, 1938-1425},
	url = {https://link.springer.com/10.1557/s43577-022-00384-5},
	doi = {10.1557/s43577-022-00384-5},
	language = {en},
	number = {6},
	urldate = {2025-01-27},
	journal = {MRS Bulletin},
	author = {Felser, Claudia and Parkin, Stuart},
	month = jun,
	year = {2022},
	note = {Number: 6},
	pages = {600--608},
}

@article{prasad_material_2022,
	title = {Material challenges for nonvolatile memory},
	volume = {10},
	issn = {2166-532X},
	url = {https://pubs.aip.org/apm/article/10/9/090401/2834976/Material-challenges-for-nonvolatile-memory},
	doi = {10.1063/5.0111671},
	language = {en},
	number = {9},
	urldate = {2025-01-19},
	journal = {APL Materials},
	author = {Prasad, Bhagwati and Parkin, Stuart and Prodromakis, Themis and Eom, Chang-Beom and Sort, Jordi and MacManus-Driscoll, J. L.},
	month = sep,
	year = {2022},
	note = {Number: 9},
	pages = {090401},
}

@article{park_magnetic_2018,
	title = {Magnetic excitations in non-collinear antiferromagnetic {Weyl} semimetal {Mn3Sn}},
	volume = {3},
	issn = {2397-4648},
	url = {https://www.nature.com/articles/s41535-018-0137-9},
	doi = {10.1038/s41535-018-0137-9},
	language = {en},
	number = {1},
	urldate = {2024-11-24},
	journal = {npj Quantum Materials},
	author = {Park, Pyeongjae and Oh, Joosung and Uhlířová, Klára and Jackson, Jerome and Deák, András and Szunyogh, László and Lee, Ki Hoon and Cho, Hwanbeom and Kim, Ha-Leem and Walker, Helen C. and Adroja, Devashibhai and Sechovský, Vladimír and Park, Je-Geun},
	month = dec,
	year = {2018},
	note = {Number: 1},
	pages = {63},
}

@article{sato_magnetic_2017,
	title = {Magnetic tunnel junctions with perpendicular easy axis at junction diameter of less than 20 nm},
	volume = {56},
	issn = {0021-4922, 1347-4065},
	url = {https://iopscience.iop.org/article/10.7567/JJAP.56.0802A6},
	doi = {10.7567/JJAP.56.0802A6},
	number = {8},
	urldate = {2025-01-15},
	journal = {Japanese Journal of Applied Physics},
	author = {Sato, Hideo and Ikeda, Shoji and Ohno, Hideo},
	month = aug,
	year = {2017},
	note = {Number: 8},
	pages = {0802A6},
}

@article{sato_perpendicular-anisotropy_2012,
	title = {Perpendicular-anisotropy {CoFeB}-{MgO} magnetic tunnel junctions with a {MgO}/{CoFeB}/{Ta}/{CoFeB}/{MgO} recording structure},
	volume = {101},
	issn = {0003-6951, 1077-3118},
	url = {https://pubs.aip.org/apl/article/101/2/022414/127813/Perpendicular-anisotropy-CoFeB-MgO-magnetic-tunnel},
	doi = {10.1063/1.4736727},
	language = {en},
	number = {2},
	urldate = {2025-01-15},
	journal = {Applied Physics Letters},
	author = {Sato, H. and Yamanouchi, M. and Ikeda, S. and Fukami, S. and Matsukura, F. and Ohno, H.},
	month = jul,
	year = {2012},
	note = {Number: 2},
	pages = {022414},
}

@article{faleev_origin_2017,
	title = {Origin of the {Tetragonal} {Ground} {State} of {Heusler} {Compounds}},
	volume = {7},
	copyright = {http://link.aps.org/licenses/aps-default-license},
	issn = {2331-7019},
	url = {https://link.aps.org/doi/10.1103/PhysRevApplied.7.034022},
	doi = {10.1103/PhysRevApplied.7.034022},
	language = {en},
	number = {3},
	urldate = {2025-01-14},
	journal = {Physical Review Applied},
	author = {Faleev, Sergey V. and Ferrante, Yari and Jeong, Jaewoo and Samant, Mahesh G. and Jones, Barbara and Parkin, Stuart S. P.},
	month = mar,
	year = {2017},
	note = {Number: 3},
	pages = {034022},
}

@article{faleev_heusler_2017,
	title = {Heusler compounds with perpendicular magnetic anisotropy and large tunneling magnetoresistance},
	volume = {1},
	copyright = {http://link.aps.org/licenses/aps-default-license},
	issn = {2475-9953},
	url = {http://link.aps.org/doi/10.1103/PhysRevMaterials.1.024402},
	doi = {10.1103/PhysRevMaterials.1.024402},
	language = {en},
	number = {2},
	urldate = {2025-01-14},
	journal = {Physical Review Materials},
	author = {Faleev, Sergey V. and Ferrante, Yari and Jeong, Jaewoo and Samant, Mahesh G. and Jones, Barbara and Parkin, Stuart S. P.},
	month = jul,
	year = {2017},
	note = {Number: 2},
	pages = {024402},
}

@article{dieny_perpendicular_2017,
	title = {Perpendicular magnetic anisotropy at transition metal/oxide interfaces and applications},
	volume = {89},
	copyright = {http://link.aps.org/licenses/aps-default-license},
	issn = {0034-6861, 1539-0756},
	url = {http://link.aps.org/doi/10.1103/RevModPhys.89.025008},
	doi = {10.1103/RevModPhys.89.025008},
	language = {en},
	number = {2},
	urldate = {2025-01-14},
	journal = {Reviews of Modern Physics},
	author = {Dieny, B. and Chshiev, M.},
	month = jun,
	year = {2017},
	note = {Number: 2},
	pages = {025008},
}

@article{sales_ferromagnetism_2014,
	title = {Ferromagnetism of {Fe3Sn} and {Alloys}},
	volume = {4},
	issn = {2045-2322},
	url = {https://www.nature.com/articles/srep07024},
	doi = {10.1038/srep07024},
	language = {en},
	number = {1},
	urldate = {2026-02-15},
	journal = {Scientific Reports},
	author = {Sales, Brian C. and Saparov, Bayrammurad and McGuire, Michael A. and Singh, David J. and Parker, David S.},
	month = nov,
	year = {2014},
	pages = {7024},
}

@article{prodan_large_2023,
	title = {Large ordered moment with strong easy-plane anisotropy and vortex-domain pattern in the kagome ferromagnet {Fe3Sn}},
	volume = {123},
	issn = {0003-6951, 1077-3118},
	url = {https://pubs.aip.org/apl/article/123/2/021901/2901979/Large-ordered-moment-with-strong-easy-plane},
	doi = {10.1063/5.0155295},
	language = {en},
	number = {2},
	urldate = {2026-02-15},
	journal = {Applied Physics Letters},
	author = {Prodan, Lilian and Evans, Donald M. and Griffin, Sinéad M. and Östlin, Andreas and Altthaler, Markus and Lysne, Erik and Filippova, Irina G. and Shova, Sergiu and Chioncel, Liviu and Tsurkan, Vladimir and Kézsmárki, István},
	month = jul,
	year = {2023},
	pages = {021901},
}

% %\section{Supplementary Figures}
% %%\begin{figure*}
%     \centering
%     \includegraphics[width=1.\linewidth]{FWHMfigsinglpeakv1.png}
%     \caption{Enter Caption}
%     \label{fig:placeholder}
% %%\begin{figure*}
%     \centering
%     \includegraphics[width=1.\linewidth]{FWHMstack500peakmerged.png}
%     \caption{Enter Caption}
%     \label{fig:placeholder}
% \end{figure*}
\end{document}